\documentclass[12pt]{article}
\usepackage{latexsym}
\usepackage{bbm}
\newcounter{saveeqn}
\newcommand{\alpheqn}%
    {\setcounter{saveeqn}{\value{equation}}%
     \stepcounter{saveeqn}%
     \setcounter{equation}{0}%
     }
\newcommand{\reseteqn}%
    {\setcounter{equation}{\value{saveeqn}}%
     }
\newtheorem{theorem}{Theorem}
\newtheorem{proposition}{Proposition}
\newtheorem{lemma}{Lemma}
\begin{document}
\title{\normalsize \hfill UWThPh-2002-4 \\[1cm] \LARGE
Dissipation in a 2-dimensional Hilbert space:\\
Various forms of complete positivity\thanks{This
research was performed within the FWF Project No.\ P14143--PHY of the
Austrian Science Foundation.}}
\author{R.A. Bertlmann and W. Grimus \\
\small Institut f\"ur Theoretische Physik, Universit\"at Wien \\
\small Boltzmanngasse 5, A--1090 Wien, Austria \\*[4.6mm] }

\date{June 2002}

\maketitle

\begin{abstract}
We consider the time evolution of the density matrix $\rho$ in a 2-dimensional
complex Hilbert space. We allow for dissipation by adding to the von
Neumann equation a term $D[\rho]$, which is of
Lindblad type in order to assure complete positivity of the time evolution.
We present five equivalent forms of $D[\rho]$.
In particular, we connect the familiar dissipation matrix $L$ with a geometric
version of $D[\rho]$, where $L$ consists of a positive sum of projectors onto
planes in $\mathbbm{R}^3$. We also study the minimal number of Lindblad terms
needed to describe the most general case of $D[\rho]$.
All proofs are worked out comprehensively, as they present at the same time a
practical procedure how to determine explicitly the different forms of
$D[\rho]$. Finally, we perform a general discussion of the asymptotic
behaviour $t \to \infty$ of the density matrix and we relate the two
types of asymptotic behaviour with our geometric version of $D[\rho]$.

\vspace*{8mm}

\end{abstract}

\newpage
\setcounter{footnote}{2}

\section{Introduction}

Particle physics is not only a field where fundamental interactions are
explored, but it has also become a testing ground for possible deviations from
the quantum-mechanical time evolution.
The time evolution of the density matrix $\rho$ is given by the master
equation
\begin{equation}\label{timeev}
\frac{d \rho}{dt} = -i H \rho + i \rho H^\dagger - D[\rho] \,,
\end{equation}
where $D[\rho]$ is a dissipative term which adds to the quantum-mechanical
term on the right-hand side of Eq.~(\ref{timeev}). Such a
term can emerge if the system under consideration is not fully closed but
interacts weakly with the environment. In general the nature of such an
interaction is unknown, but experimental data from suitable systems can be
used to place bounds on the parameters of such hypothesized interactions. In
the general case, the Weisskopf--Wigner approximation \cite{WWA}
allows to incorporate also unstable particles, by
using non-hermitian Hamiltonians $H = M - i \Gamma/2$, where $M$ and
$\Gamma \geq 0$ are hermitian.

It is not only necessary that the
time evolution (\ref{timeev}) respects $\rho(t) \geq 0$ $\forall \, t$,
but the stronger requirement of \emph{complete positivity} \cite{stinespring}
seems to be a natural and physical concept (for the general structure of
completely positive maps, see Refs.~\cite{choi,fujiwara}).
This concept is defined in the
following way. Let us assume that the system under consideration is described
by elements of the finite complex Hilbert space $\mathcal{H}$ 
($\dim \mathcal{H} \equiv d < \infty$) with time evolution
$\rho(t) \equiv \gamma_t (\rho)$ with $\rho(0) = \rho$.
Considering in addition
the finite-dimensional Hilbert space $\mathbbm{C}^n$, one can extend the time
evolution $\gamma_t$ on $\mathcal{H}$ to a time evolution $\gamma_{n;t}$
on $\mathcal{H} \otimes \mathbbm{C}^n $ by defining
$\gamma_{n;t} = \gamma_t \otimes \mathbbm{1}$. If the time evolution
$\gamma_t$ derived from Eq.~(\ref{timeev}) with the dissipative term
$D[\rho]$ has the property that $\gamma_{n;t}(\rho) \geq 0$ is valid for all
times $t$, all $n = 0, 1, 2, \ldots$ and all density
matrices on the space $\mathcal{H} \otimes \mathbbm{C}^n $, then
$\gamma_t$ is called completely positive.

Complete positivity of $\gamma_t$ determines the general structure of
$D[\rho]$ \cite{lindblad,goriniKS} (see also
Refs.~\cite{goriniFVKS,goriniK,alickiL,holevo}).
It has been shown by Lindblad \cite{lindblad} (see also Ref.~\cite{adler})
that $\gamma_t$ of Eq.~(\ref{timeev})
is completely positive if and only if $D[\rho]$ has the structure
\begin{eqnarray}\label{diss}
D[\rho] &=& \frac{1}{2}
\sum_{j=1}^r \left( A_j^\dagger A_j \rho + \rho A_j^\dagger A_j
- 2\, A_j \rho A_j^\dagger \right) \nonumber \\
&=& \frac{1}{2}
\sum_{j=1}^r \left( [ A_j^\dagger, A_j \rho ] +
[ \rho A_j^\dagger, A_j ] \right) \, ,
\end{eqnarray}
where the operators $A_j$ act on $\mathcal{H}$.

Using the relation
\begin{equation}
A_j = B_j + s_j \mathbbm{1}
\quad \mathrm{with} \quad \mathrm{Tr}\, B_j = 0 \,,
\end{equation}
the dissipative term (\ref{diss}) can be reformulated as 
\begin{equation}\label{diss1}
D[\rho] = -i [ \Delta H, \rho ] + D'[\rho] 
\quad \mathrm{with} \quad
\Delta H = \frac{i}{2} \sum_{j=1}^r \left( s_j B_j^\dagger - s_j^* B_j \right) \,,
\end{equation}
where $D'[\rho]$ is obtained from $D[\rho]$ by the replacement
$A_j \to B_j$. This reformulation has the effect that part of $D[\rho]$ is
shifted into the quantum-mechanical term of the time evolution (\ref{timeev}),
such that a new Hamiltonian $H' = H + \Delta H$ appears. Note that for
hermitian operators $A_j$ we have $\Delta H = 0$. In the space of traceless
operators on the Hilbert space $\mathcal{H}$ we can choose a basis
$\left\{ F_j |\, j = 1, \ldots , d^2 -1 \right\}$ with the property
\begin{equation}
\mathrm{Tr}\, \left( F_j^\dagger F_k \right) = \delta_{jk} 
\end{equation}
and expand the operators $B_j$ as
\begin{equation}
B_j = \sum_{k=1}^{d^2-1} C_{kj} F_k \,.
\end{equation}
Then we obtain the expression
\begin{equation}\label{diss2}
D'[\rho] = -\frac{1}{2} \sum_{j=1}^r \left( \,
[B_j, \rho B_j^\dagger] + [B_j \rho, B_j^\dagger] \, \right) = 
-\frac{1}{2} \sum_{k,l=1}^{d^2-1} c_{kl} \left( \,
[F_k, \rho F_l^\dagger] + [F_k \rho, F_l^\dagger] \, \right) \,,
\end{equation}
where $(c_{kl})$ is a positive matrix defined by
\begin{equation}\label{cc}
c_{kl} = \sum_{j=1}^r C_{kj} C_{lj}^* \,.
\end{equation}
The form (\ref{diss2}) of the dissipative term has been derived 
by Gorini, Kossakowski and Sudarshan \cite{goriniKS}.
It is equivalent to the Lindblad form (\ref{diss}). Thus the
time evolution with $H$ and $D[\rho]$ is equivalent to the one with $H'$ and
$D'[\rho]$. 

In particle physics, searches for
deviations from the quantum-mechanical time evolution are going on in
neutral meson-antimeson systems ($K^0 \bar K^0$ and $B^0 \bar B^0$)
(for a list of papers see, e.g.,
Refs.~\cite{ellisHNS,ellisLMN,banksSP,huetP,benatti96,benatti97,benatti98,%
andrianov,BG01,benatti01})
and neutrino physics (for a list of papers see, e.g.,
Refs.~\cite{enqvist,chang,lisi,klapdor,benatti00}).
The importance of complete positivity, in particular, in
$K^0 \bar K^0$ and analogous systems, has been stressed in
Refs.~\cite{benatti96,benatti98}.
For instance, only if $\gamma_t$ is completely positive, the positivity of the
time evolution $\gamma_t \otimes \gamma_t$ in the space
$\mathcal{H} \otimes \mathcal{H}$ is guaranteed. This follows from the
decomposition
$\gamma_t \otimes \gamma_t =
(\gamma_t \otimes \mathbbm{1})\,(\mathbbm{1} \otimes \gamma_t)$,
where both factors are positive according to complete positivity.
Forming the tensor product $\gamma_t \otimes \gamma_t$ is a method for
implementing the
1-particle time evolution at the 2-particle level, which is an often used
procedure, e.g., in the $K^0 \bar K^0$ system. For an example where $\gamma_t$
is only positive but not completely positive, with ensuing non-positivity of
$\gamma_t \otimes \gamma_t$, see Ref.~\cite{benatti98}.

For simplicity we assume a hermitian
Hamiltonian $H$ from now on, but this assumption is
irrelevant for our discussion of complete positivity; it only concerns the
investigation of the asymptotic limit of the time evolution, where loss of
probability due to a non-hermitian $H$ as obtained in the Weisskopf--Wigner
approximation \cite{WWA} will lead to $\rho(t) \to 0$
for $t \to \infty$.

Apart from complete positivity of the time evolution,
the assumptions on which the present work is based are the following:
\begin{enumerate}
\item
We work in a 2-dimensional complex Hilbert space, which we can identify with
$\mathbbm{C}^2$;
\item
we assume hermitian Lindblad operators, i.e.,
$A_j = A_j^\dagger$\hspace{2pt} $\forall j$.
\end{enumerate}
Both assumptions are crucial for the following discussions. The first one is
motivated by the applications in particle physics, whereas the
second assumption guarantees that the entropy
$S[\rho] = - \mbox{Tr}\, (\rho \ln \rho)$
cannot decrease as a function of time \cite{narnhofer}.
In this framework we will discuss the different forms of the dissipative
term $D[\rho]$ used in the literature and we will show their
equivalence. We will put emphasis on the formulation of the time
evolution (\ref{timeev}) in $\mathbbm{R}^3$, where we will represent
$D[\rho]$ as a positive sum over projectors onto planes. We will also
study in detail the matrix formulation of $D[\rho]$ as advocated by
Benatti and Floreanini, e.g., in Refs.~\cite{benatti96,benatti98}, and
relate this formulation with the geometric version of $D[\rho]$ as a sum
of projectors. Finally, we will investigate the limit $t \to \infty$ of
the density matrix.

\section{Equivalent forms of the dissipative term}

We start with the original form of the dissipative term (\ref{diss}) and take
into account that we confine ourselves to hermitian Lindblad operators
$A_j$. Thus $D[\rho]$ simplifies to
\begin{equation}\label{A}
\mbox{Form A:} \quad
D[ \rho ] = \frac{1}{2} \sum_{j=1}^r
\left( A_j^2 \rho + \rho A_j^2 - 2\, A_j \rho A_j \right) =
\frac{1}{2} \sum_{j=1}^r [ A_j, [ A_j, \rho]\,] \,.
\end{equation}
The number of Lindblad terms in Eq.~(\ref{A}) is denoted by $r$.

Next we note that the dissipative term can be rewritten in terms of projectors
$P_j$ and their orthogonal complements $P^\bot_j = \mathbbm{1} - P_j$ as
\begin{equation}\label{B}
\mbox{Form B:} \quad
D[ \rho ] = \frac{1}{2} \sum_{j=1}^r  \lambda_j
\left( P_j \rho P^\bot_j + P^\bot_j \rho P_j \right) \,.
\end{equation}
The projectors $P_j$ are non-trivial projectors in $\mathbbm{C}^2$, which can
be parameterized as
\begin{equation}\label{Pj}
P_j = \frac{1}{2} \left( \mathbbm{1} + \vec{n}_j \cdot \vec{\sigma} \right) \,,
\end{equation}
where the $\vec{n}_j$ are real unit vectors and $\vec{\sigma}$ denotes the
vector of Pauli matrices. The quantities $\lambda_j$ are real, positive
numbers.
\begin{proposition}
Forms A and B of $D[\rho]$ are equivalent.
\end{proposition}
\textbf{Proof:} A general hermitian $2 \times 2$ matrix $A_j$
can be represented by
\begin{equation}\label{Aj}
A_j = \frac{1}{2} \left( a_j \mathbbm{1} +
\sqrt{\lambda_j} \vec{n}_j \cdot \vec{\sigma} \right) \,,
\end{equation}
where $a_j$ and $\lambda_j$ are real numbers ($\lambda_j \geq 0$) and
$\vec{n}_j$ is a unit vector.
Note that the part of $A_j$ proportional to the unit matrix does not
contribute to $D[\rho]$, as evident from Eq.~(\ref{A}). Therefore, $\lambda_j$
must not be zero in order to have a non-trivial effect of $A_j$. Consequently,
we are allowed to replace $A_j$ by $\sqrt{\lambda_j} P_j$ in
$D[\rho]$. Using Eq.~(\ref{A}) and $P_j^2 = P_j$, we derive
\begin{equation}
D[ \rho ] = \frac{1}{2}
\sum_{j=1}^r \lambda_j \left( P_j \rho + \rho P_j - 2\, P_j \rho P_j \right)
\,,
\end{equation}
which can be rewritten in Form B of Eq.~(\ref{B}). $\Box$

\vspace{4mm}

The time evolution (\ref{timeev}) is easily reformulated as a differential
equation for a real 3-vector $\vec{\rho}$ by using
\begin{equation}
\rho = \frac{1}{2} \left( \mathbbm{1} + \vec{\rho} \cdot \vec{\sigma} \right)
\quad \mathrm{and} \quad
H = \frac{1}{2} \left( \mathbbm{1} + \vec{h} \cdot \vec{\sigma} \right) \,.
\end{equation}
Since for simplicity we have assumed that $H$ is hermitian, the 3-vector
$\vec{h}$ is real as well (for an extension to non-hermitian
Hamiltonians see, e.g., Ref.~\cite{BH}).
The new version of Eq.~(\ref{timeev}) is given by
\begin{equation}\label{timeev1}
\frac{\mathrm{d}\vec{\rho}}{\mathrm{d}t} =
\vec{h} \times \vec{\rho} - L \vec{\rho} \,.
\end{equation}
With Form A or B of the dissipative term and Eq.~(\ref{Aj}) or (\ref{Pj}), 
we arrive at a geometric version of $D[\rho]$
\begin{equation}\label{C}
\mbox{Form C:} \quad
L \vec{\rho} =
\frac{1}{2} \sum_{j=1}^r \lambda_j \left(
\vec{\rho} - \vec{n}_j\, \vec{n}_j \cdot \vec{\rho} \right) \,,
\end{equation}
where the matrix $L$ is a positive linear combination of projectors
\begin{equation}\label{P}
\mathcal{P}(\vec{n}_j) = \mathbbm{1}_3 - \vec{n}_j \vec{n}_j^T \,.
\end{equation}
The projector (\ref{P})
projects onto the plane orthogonal to $\vec{n}_j$.
Clearly, Form C is equivalent to the the previous forms of the
dissipative term.

The dissipation matrix $L$ in Form C can be reformulated as \cite{benatti96}
\begin{equation}\label{D}
\mbox{Form D:} \quad
L_{\alpha \beta} = \frac{1}{2} \left(
\Lambda\, \delta_{\alpha \beta} -
\mathbf{q}_\alpha \cdot \mathbf{q}_\beta \right)
\end{equation}
with
\begin{equation}
\mathbf{q}_\alpha \in \mathbbm{R}^r \quad (\alpha = 1,2,3)
\quad \mathrm{and} \quad
\Lambda = \sum_{\alpha = 1}^3 | \mathbf{q}_\alpha |^2 \,.
\end{equation}
\begin{proposition}
Forms C and D are equivalent.
\end{proposition}
\textbf{Proof:} The proof of this statement amounts to a mere rewriting of the
elements of $L$ by
\begin{equation}
\sum_{j=1}^r \lambda_j \left( \delta_{\alpha\beta} -
(\vec{n}_j)_\alpha (\vec{n}_j)_\beta \right) =
\sum_{j=1}^r \lambda_j \delta_{\alpha\beta} -
\sum_{j=1}^r \sqrt{\lambda_j}
(\vec{n}_j)_\alpha \sqrt{\lambda_j} (\vec{n}_j)_\beta \,.
\end{equation}
Then we define
\begin{equation}\label{q}
\mathbf{q}_\alpha =
\left( \begin{array}{c}
\sqrt{\lambda_1}\, (\vec{n}_1)_\alpha \\ \vdots \\
\sqrt{\lambda_r}\, (\vec{n}_r)_\alpha
\end{array} \right)
\quad \mathrm{and} \quad
\Lambda = \sum_{j=1}^r \lambda_j \,.
\end{equation}
The sum over the square of the lengths of the three vectors
$\mathbf{q}_\alpha$ is performed via
\begin{equation}
\sum_{\alpha = 1}^3 | \mathbf{q}_\alpha |^2 =
\sum_{\alpha = 1}^3 \sum_{j = 1}^r \lambda_j (\vec{n}_j)_\alpha^2 =
\sum_{j = 1}^r \lambda_j \left( \sum_{\alpha = 1}^3
(\vec{n}_j)_\alpha^2 \right) =
\sum_{j = 1}^r \lambda_j \,.
\end{equation}
Here, we have used that the $\vec{n}_j$ are 3-dimensional unit vectors.
Thus we have obtained Form D of $L$ from Form C. This procedure can be
reversed: given three vectors $\mathbf{q}_\alpha \in \mathbbm{R}^r$, we can use
Eq.~(\ref{q}) to construct $r$ unit vectors $\vec{n}_j$ and positive numbers
$\lambda_j$. In this way, we gain Form C from Form D.
$\Box$

\vspace{4mm}

The question arises how many terms are necessary in $D[\rho]$ in the most
general case. This question is answered by the following theorem.
\begin{theorem}\label{l}
The most general case is covered by three Lindblad terms in $D[\rho]$, i.e.,
if $D[\rho]$ is given by a sum over more than three terms, then it
can be rewritten as a sum of at most three terms. If $D[\rho]$ is formulated
with the minimal number of terms, then, using Forms B or C,
there are three distinct minimal cases referring to one, two or
three linearly independent vectors $\vec{n}_j$ with
$\lambda_j > 0$;
we will denote these cases by an
\emph{index} $\ell \in \{ 1,2,3 \}$.\footnote{Obviously, in the
minimal formulation of $D[\rho]$ we have $r \equiv \ell$.}
\end{theorem}
\textbf{Proof:} Let us first assume that $r > 3$.
Since there are only three vectors $\mathbf{q}_\alpha$, we can find a
rotation $R$ acting on $\mathbbm{R}^r$ such that
\begin{equation}\label{R}
R \mathbf{q}_\alpha = \left( \begin{array}{c}
\mathbf{Q}_\alpha \\ 0 \\ \vdots \\ 0
\end{array} \right)
\quad \mathrm{with} \quad \mathbf{Q}_\alpha \in \mathbbm{R}^3 \,.
\end{equation}
In the most general case the set of vectors
$Q = \{ \mathbf{Q}_\alpha | \alpha = 1,2,3 \}$
is linearly independent and can be parameterized by
\begin{equation}
\mathbf{Q}_\alpha =
\left( \begin{array}{c} q_\alpha^1 \\ q_\alpha^2 \\ q_\alpha^3
\end{array} \right) =
\left( \begin{array}{c}
\sqrt{\mu_1}\, (\vec{m}_1)_\alpha \\
\sqrt{\mu_2}\, (\vec{m}_2)_\alpha \\
\sqrt{\mu_3}\, (\vec{m}_3)_\alpha
\end{array} \right) \,.
\end{equation}
The real and positive numbers $\mu_j$ are chosen in such a way that the vector
$\vec{m}_1$, extracted from the first elements of the
vectors $\mathbf{Q}_\alpha$, is a unit vector;
the same is done for the second and third
elements.\footnote{This procedure is analogous to the one to prove that Form C
follows from Form D.} In this case we have an index $\ell = 3$. If $Q$ spans a
2-dimensional space, we choose the rotation $R$ such that the third elements
of all $\mathbf{Q}_\alpha$ are zero; if $Q$ spans a 1-dimensional space, the
second and third elements are taken to be zero. These two cases refer to index
$\ell = 2$ and 1, respectively. Form D of the dissipative term tells us that
$D[\rho]$ is independent of any rotation $R$. Consequently, we arrive at
\begin{equation}
\sum_{j=1}^r \lambda_j \left( \delta_{\alpha\beta} -
(\vec{n}_j)_\alpha (\vec{n}_j)_\beta \right) =
\sum_{j=1}^\ell \mu_j \left( \delta_{\alpha\beta} -
(\vec{m}_j)_\alpha (\vec{m}_j)_\beta \right) \,,
\end{equation}
which proves the theorem for $r>3$. For $r \leq 3$ we follow analogous steps
performed after Eq.~(\ref{R}), with the rotation $R$ acting now on a 3 or
2-dimensional space. For $r=1$ the procedure is trivial.
$\Box$

\vspace{4mm}

On the other hand, to find the minimal number of operators $A_j$
needed for the most
general dissipative term (\ref{diss}), we could start 
from the Gorini-Kossakowski-Sudarshan expression (\ref{diss2}).
Transforming back to the Lindblad structure
(\ref{diss}) by using the relations
\begin{equation}
c_{kl} = \left( U \, \hat c \, U^\dagger \right)_{kl} \quad \mathrm{and} \quad
B_j = \sum_{k=1}^{d^2-1} U_{kj} \sqrt{\hat c_{jj}} \, F_k \,,
\end{equation}
where $\hat c \geq 0$ is diagonal and $U$ a unitary matrix,
it can be seen that, in the general case, we obtain $d^2-1$ terms
in $D[\rho]$ of Eq.~(\ref{diss}); if some of the elements 
$\hat c_{jj}$ are zero, we will have less than $d^2-1$ terms. Applying
this to $d = 2$, we find at most three terms, which agrees with 
the result of the explicit calculations leading to Theorem \ref{l}.

\section{Complete-positivity conditions on the dissipation matrix $L$}

Benatti and Floreanini \cite{benatti96,benatti97,benatti98}
parameterized the dissipation matrix $L$ by 6 real constants and expressed
complete positivity in the form of
inequalities satisfied by these parameters. Thus they have the version
\begin{equation}
\mbox{Form E:} \quad
L = 2 \left(
\begin{array}{ccc}
a & b & c \\ b & \alpha & \beta \\ c & \beta & \gamma
\end{array} \right) \,,
\end{equation}
together with
\alpheqn
\begin{eqnarray}
&
\begin{array}{c}
2R \equiv \alpha + \gamma - a \geq 0 \,, \\
2S \equiv a + \gamma - \alpha \geq 0 \,, \\
2T \equiv a + \alpha - \gamma \geq 0 \,;
\end{array}
& \label{a} \\ &
RS \geq b^2\,, \quad RT \geq c^2 \,, \quad ST \geq \beta^2 \,;
&  \label{b} \\ &
RST \geq 2 b c \beta + R \beta^2 + S c^2 + T b^2 \,.
& \label{c}
\end{eqnarray}
\reseteqn
We will show now that Form E is equivalent to the forms presented in
the previous section.

First we want to put Eqs.~(\ref{a}), (\ref{b}), (\ref{c}) into a simpler
equivalent form.
\begin{lemma}\label{LM}
Given $L$, there exists a symmetric matrix $M$ such that
\begin{equation}\label{M}
L = \frac{1}{2} (\mathrm{Tr}\, M\, \mathbbm{1}_3 - M) \,.
\end{equation}
In terms of $M$, Eqs.~(\ref{a}), (\ref{b}), (\ref{c}) are given by
\begin{equation}\label{condM}
\begin{array}{rcl}
\mbox{(i)} & M_{\alpha\alpha} \geq 0 & (\alpha = 1,2,3 ) \,, \\
\mbox{(ii)} & M_{\alpha\alpha} M_{\beta\beta} \geq M_{\alpha\beta}^2 &
\,\forall\, \alpha \neq \beta \,, \\
\mbox{(iii)} & \det M \geq 0 \,, &
\end{array}
\end{equation}
respectively.
\end{lemma}
\textbf{Proof:} For $\alpha \neq \beta$ we have
$M_{\alpha\beta} = -2 L_{\alpha\beta}$. The diagonal elements of $M$ are
determined by
$M_{\alpha\alpha} = -L_{\alpha\alpha} + L_{\beta\beta} +
L_{\gamma\gamma}$, where $\alpha \neq \beta \neq \gamma \neq \alpha$.
The second part of the lemma is
proven by plugging Eq.~(\ref{M}) into Eqs.~(\ref{a}), (\ref{b}), (\ref{c}).
$\Box$

\vspace{4mm}

The next lemma will allow us to connect $M$ possessing properties
(\ref{condM}) with Form D of the dissipative term.
\begin{lemma}\label{Mqq}
A real and symmetric $3 \times 3$ matrix
$M$ has the properties of Eq.~(\ref{condM})
if and only if there exist vectors $\mathbf{q}_\alpha \in \mathbbm{R}^3$
($\alpha = 1,2,3$) such that
\begin{equation}\label{product}
M_{\alpha\beta} = \mathbf{q}_\alpha \cdot \mathbf{q}_\beta \,.
\end{equation}
\end{lemma}
\textbf{Proof:} \\
($\Leftarrow$) This direction of the proof is quickly dealt with.
Since
$M_{\alpha\alpha} = \mathbf{q}_\alpha^2 \geq 0$, property (i) is
valid. Furthermore, for $\alpha \neq \beta$, with the Cauchy--Schwarz
inequality we derive
$M_{\alpha\beta}^2 = (\mathbf{q}_\alpha \cdot \mathbf{q}_\beta)^2 \leq
\mathbf{q}_\alpha^2\, \mathbf{q}_\beta^2 = M_{\alpha\alpha} M_{\beta\beta}$
and property (ii) holds as well. Defining a $3 \times 3$ matrix
$q = ( \mathbf{q}_1, \mathbf{q}_2, \mathbf{q}_3)$, we have
$\det M = \det (q^T q) = (\det q)^2 \geq 0$ and thus property
(iii). This completes the first half of the proof.\\
($\Rightarrow$) This direction of the proof is more involved.
If all $M_{\alpha\alpha}$ were zero, then according to property (ii) of
Eq.~(\ref{condM}) we would have $M = 0$. Therefore, at least one of the
diagonal elements of $M$ is non-zero. Without loss of generality we assume
$M_{11} > 0$. First we consider the case
\begin{equation}\label{cond1}
M_{11}M_{22} - M_{12}^2 > 0 \,.
\end{equation}
Denoting the elements of $\mathbf{q}_\alpha$ by $q_\alpha^j$,
we can define a vector $\mathbf{q}_1$ by
$q_1^1 = \sqrt{M_{11}}$, $q_1^2 = q_1^3 = 0$.
From $M_{12} = \mathbf{q}_1 \cdot \mathbf{q}_2$, it follows after a sign
choice that
$q_2^1 = M_{12}/\sqrt{M_{11}}$. Taking into account that
$\mathbf{q}_2 \cdot \mathbf{q}_2 = M_{22}$ and defining $q_2^3 = 0$,
the first two vectors are given by
\begin{equation}\label{q12}
\mathbf{q}_1 = \left( \begin{array}{c} \sqrt{M_{11}} \\ 0 \\ 0
\end{array} \right) \,, \quad
\mathbf{q}_2 = \left( \begin{array}{c} M_{12}/\sqrt{M_{11}} \\
\sqrt{M_{22} - M_{12}^2/M_{11}} \\ 0
\end{array} \right) \,.
\end{equation}
With the relation of Eq.~(\ref{cond1}) we find that $\mathbf{q}_2$ is a
well-defined real 3-vector. Next we use
$M_{13} = \mathbf{q}_1 \cdot \mathbf{q}_3$ and
$M_{23} = \mathbf{q}_2 \cdot \mathbf{q}_3$ and obtain
\begin{equation}
q_3^1 = \frac{M_{13}}{\sqrt{M_{11}}}
\quad \mbox{and} \quad
q_3^2 = \frac{M_{23} - M_{12} M_{13}/M_{11}}{\sqrt{M_{22} -
M_{12}^2/M_{11}}} \,.
\end{equation}
It remains to take into account
$\mathbf{q}_3 \cdot \mathbf{q}_3 = M_{33}$.
After some algebra we arrive at
\begin{equation}
(q_3^3)^2 =
M_{33} - (q_3^1)^2 - (q_3^2)^2 =
\frac{\det M}{M_{11} M_{22} - M_{12}^2} \geq 0 \,.
\end{equation}
The positivity follows from property (iii) of Eq.~(\ref{condM}). Thus, 
$q_3^3$ is well-defined and we have
proven relation (\ref{product}), provided condition (\ref{cond1}) holds.

It remains to check the same for the special case
\begin{equation}\label{cond2}
M_{11} M_{22} = M_{12}^2 \,,
\end{equation}
which was excluded by Eq.~(\ref{cond1}).
From Eq.~(\ref{cond2}) we obtain
$M_{12} = \eta \sqrt{M_{11} M_{22}}$ with $\eta = \pm 1$ and
\begin{equation}
\det M = -\left(\sqrt{M_{11}} M_{23} - \eta \sqrt{M_{22}} M_{13}
\right)^2 \,.
\end{equation}
Since $\det M \geq 0$ (see Eq.~(\ref{condM})), it follows that
\begin{equation}
M_{23} = \eta M_{13} \sqrt{M_{22}/M_{11}} \,.
\end{equation}
With this relation and taking into account condition (\ref{cond2}),
it is easy to check that
\begin{equation}
\mathbf{q}_1 = \sqrt{M_{11}} \left(
\begin{array}{c} 1 \\ 0 \\ 0
\end{array} \right) \,, \quad
\mathbf{q}_2 = \eta \sqrt{M_{22}} \left(
\begin{array}{c} 1 \\ 0 \\ 0
\end{array} \right) \,, \quad
\mathbf{q}_3 = \left( \begin{array}{c}
M_{13}/\sqrt{M_{11}} \\ \sqrt{M_{33} - M_{13}^2/M_{11}} \\ 0
\end{array} \right)
\end{equation}
represents a consistent choice of vectors, which fulfills equation
(\ref{product}). This completes the proof. $\Box$

\vspace{4mm}

With Lemmata \ref{LM} and \ref{Mqq} we now readily see that
a matrix $L$ fulfills the conditions of
Eqs.~(\ref{a}), (\ref{b}), (\ref{c}) if and only if $L$ is given by
$L_{\alpha\beta} = \frac{1}{2} (\Lambda\, \delta_{\alpha\beta} -
\mathbf{q}_\alpha \cdot \mathbf{q}_\beta )$, where
$\mathbf{q}_\alpha \in \mathbbm{R}^3$ and
$\Lambda = \sum_{\alpha=1}^3 | \mathbf{q}_\alpha |^2$.
\begin{proposition}
Forms D and E of the dissipation matrix $L$ are equivalent.
\end{proposition}
\textbf{Proof:} According to Lemma \ref{LM}, from the matrix $L$ we construct
a matrix $M$ with properties (\ref{condM}). Lemma \ref{Mqq} tells us that such
an $M$ is represented by a matrix of scalar products (see Eq.~(\ref{product}))
and vice versa. Therefore, our statement is true. $\Box$

\section{The asymptotic limit $t \to \infty$}

Now we consider the asymptotic limit of the density matrix $\rho(t)$ with time
evolution (\ref{timeev}) \cite{BG01}. For this purpose we use the
results of Theorem \ref{l}, where we have also defined the index
$\ell$ of $L$. For a general study of the large time behaviour of the density
matrix starting with expression (\ref{diss2}) see Ref.~\cite{lendi}.
\begin{theorem}\label{l=2,3}
For index $\ell=2$ or $\ell=3$,
the asymptotic limit of the density matrix is given by
\begin{equation}
\lim_{t \to \infty} \rho(t) = {\textstyle \frac{1}{2}} \mathbbm{1} \,.
\end{equation}
\end{theorem}
\textbf{Proof:}
This statement is most easily proven by using Form C,
Eq.~(\ref{C}), of the dissipative term and Eq.~(\ref{timeev1}) of the time
evolution. Denoting by $\mathcal{A}$ the operator on the right-hand side of
Eq.~(\ref{timeev1}), we have
\begin{equation}\label{AA}
\mathcal{A} \mathbf{x} = \vec{h} \times \mathbf{x} -
\frac{1}{2} \sum_{j=1}^\ell \lambda_j \mathcal{P}(\vec{n}_j) \mathbf{x}
\end{equation}
for arbitrary complex 3-vectors $\mathbf{x}$. If we can show that every
eigenvalue $c$ of $\mathcal{A}$ fulfills $\mathrm{Re}\, c < 0$, the theorem is
proven. Let $\mathbf{x}$ be a normalized eigenvector with eigenvalue
$c$. Then, with $\mathbf{x}^\dagger \mathbf{x} = 1$ we have
\begin{equation}
c = \mathbf{x}^\dagger \mathcal{A} \mathbf{x} = -2i \vec{h} \cdot
( \mathrm{Re}\, \mathbf{x} \times \mathrm{Im}\, \mathbf{x})
-\frac{1}{2} \sum_{j=1}^\ell \lambda_j
\mathbf{x}^\dagger \mathcal{P}(\vec{n}_j) \mathbf{x}
\,,
\end{equation}
and, consequently,
\begin{equation}
\mathrm{Re}\, c =
-\frac{1}{2} \sum_{j=1}^\ell \lambda_j
\mathbf{x}^\dagger \mathcal{P}(\vec{n}_j) \mathbf{x}
\,.
\end{equation}
Since the projectors $\mathcal{P}(\vec{n}_j)$ are positive operators and
$\lambda_j >0$, we have $\mathrm{Re}\, c \leq 0$. If $\mathrm{Re}\, c = 0$, it
is necessary that
$\mathbf{x}^\dagger \mathcal{P}(\vec{n}_j) \mathbf{x} = 1 - | \vec{n}_j \cdot
\mathbf{x} |^2 = 0$ $\forall j = 1, \ldots, \ell$. Therefore, the eigenvector
$\mathbf{x}$ is proportional to $\vec{n}_j$ $\forall j = 1, \ldots,
\ell$. This is a contradiction for $\ell = 2$ or 3 independent vectors
$\vec{n}_j$ and, indeed, every eigenvalue $c$ has a negative non-zero real
part. $\Box$

\begin{theorem}\label{l=1}
For $\ell=1$ ($\vec{n}_1 \equiv \vec n$, $P_1 \equiv P$, 
$\lambda_1 \equiv \lambda$) we have either
$\vec h \parallel \vec n$ or, equivalently, $[H, P] = 0$, in which case we
obtain
\begin{equation}\label{lim2}
\lim_{t \to \infty} \rho(t) = 
P \rho(0) P + P^\bot \rho(0) P^\bot \,;
\end{equation}
or $\vec h \not\hskip1.5pt\parallel \vec n$, i.e.,
$[H, P] \neq 0$, then
the asymptotic limit of $\rho(t)$ is the same as in Theorem \ref{l=2,3}.
\end{theorem}
\textbf{Proof:}
We follow the same strategy as in the proof of the previous theorem. Thus
either $\mathrm{Re}\, c < 0$ or the eigenvector $\mathbf{x}$ is proportional
to $\vec{n}$, in which case $\mathrm{Re}\, c = 0$. Assuming
$\mathbf{x} = \vec{n}$ without loss of generality, we have now 
$
\mathcal{A} \mathbf{x} = \vec{h} \times \vec{n} = c \vec{n} 
$,
where the eigenvalue $c$ is imaginary. This equation is only soluble for
$\vec{h} \propto \vec{n}$, whence it follows that $c=0$. The
relation $\vec{h} \propto \vec{n}$
is equivalent to $[H, P] = 0$. In this case we can write the
Hamiltonian as $H = h P + h' P^\bot$. Decomposing an arbitrary
density matrix $\rho$ as
\begin{equation}
\rho = \rho_0 + \rho_1
\quad \mathrm{with} \quad
\rho_0 = P \rho P + P^\bot \rho P^\bot 
\quad \mathrm{and} \quad
\rho_1 = P \rho P^\bot + P^\bot \rho P \,,
\end{equation}
we find 
\begin{equation}
\frac{d\rho_0}{dt} = 0 
\quad \mbox{and} \quad
\frac{d\rho_1}{dt} = [-i(h - h')-\lambda/2]\, P \rho P^\bot + 
                     [ i(h - h')-\lambda/2]\, P^\bot \rho P \,.
\end{equation}
Therefore, we arrive at
\begin{equation}
\rho_0(t) = \rho_0(0)
\quad \mbox{and} \quad
\rho_1(t) = e^{[-i(h-h')-\lambda/2]t}\, P \rho(0) P^\bot + 
e^{[i(h-h')-\lambda/2]t}\, P^\bot \rho(0) P \,.
\end{equation}
Since $\lim_{t \to \infty} \rho_1(t) = 0$, the theorem is proven.
$\Box$

\vspace{4mm}

The different limits of $\rho(t)$ discussed in Theorems \ref{l=2,3} and
\ref{l=1} have been noticed in Refs.~\cite{BG01,chang,benatti00}.
In the case of $\ell = 1$ and $[H, P] = 0$, 
the limit (\ref{lim2}) of the density matrix has the form
$\rho = \mu P + (1-\mu) P^\bot$ with $0 \leq \mu \leq 1$; 
all density matrices obeying $d\rho/dt = 0$ have this form for $[H,P] = 0$.
For $\ell = 1$ and $[H,P] \neq 0$, and for $\ell = 2, 3$, the unique
density matrix which is time-independent is proportional to
the unit matrix.

\section{Summary and Conclusions}

In this paper we have considered the quantum-mechanical time evolution of
a $2 \times 2$ density matrix $\rho(t)$, where the von Neumann equation is
modified by a dissipative term $D[\rho]$, which, therefore, must be of
the Lindblad type or, equivalently, of the Gorini-Kossakowski-Sudarshan type
in order to assure a completely positive time evolution.
We have, furthermore, assumed that the Lindblad operators
$A_j$ are all hermitian which ensues that the entropy is non-decreasing with
time. Our starting point conforms with many applications of the open-systems
approach in particle physics.

We have discussed five equivalent forms of
$D[\rho]$ which all have their merits depending on the problem considered. We
have put particular emphasis on the time evolution in the form of
Eq.~(\ref{timeev1}), where the density matrix and the Hamiltonian are
represented by real 3-vectors, and Form
C, Eq.~(\ref{C}), of the dissipative term, where the dissipative term
is a positive linear 
combination of projectors onto 2-dimensional planes in $\mathbbm{R}^3$.

We have studied the question of the minimal number $\ell$ of Lindblad terms
needed in order to reproduce a given $D[\rho]$
and formulated the result in Theorem \ref{l};
the proof of this theorem represents at the same time a procedure how to
determine $\ell$ in practice. An other procedure would be to start with the
Gorini-Kossakowski-Sudarshan expression from which it also can be seen that
$D[\rho]$ can be generally decomposed into 3 terms.

We have also connected the approach
where the dissipative term is given by a matrix $L$ specified by the conditions
(\ref{a}), (\ref{b}), (\ref{c}) with the geometric picture of $D[\rho]$ as
given by Form C, Eq.~(\ref{C}). Again, the proof which shows the equivalence
between the two approaches, given by Lemmata \ref{LM} and \ref{Mqq}, indicates
a practical way to obtain the projectors $\mathcal{P}(\vec{n}_j)$ (\ref{P})
associated with $D[\rho]$.

Finally, we have presented a general discussion of the limit $t \to
\infty$ of $\rho(t)$, where the usefulness of Form C of $D[\rho]$ was
exemplified.

\end{document}